\documentclass{article}
\usepackage{graphicx}
\usepackage{amsmath}

\begin{document}

\title{Linking Maxwell, Helmholtz and Gauss through the Linking Integral}
\author{F. De\ Zela \\
Departamento de Ciencias, Secci\'{o}n F\'{i}sica, \\
Pontificia Universidad Cat\'{o}lica del Per\'{u}, Ap.1761, Lima,Per\'{u}.\\
e-mail:fdezela@fisica.pucp.edu.pe}
\maketitle

\begin{abstract}
We take the Gauss' linking integral of two curves as a starting point to
discuss the connection between the equation of continuity and the
inhomogeneous Maxwell equations. Gauss' formula has been discussed before,
as being derivable from the line integral of a magnetic field generated by a
steady current flowing through a loop. We argue that a purely geometrical
result - such as Gauss' formula - cannot be claimed to be derivable from a
law of Nature, i.e., from one of Maxwell's equations, which is the departing
point for the calculation of the magnetic field. We thus discuss anew the
derivation of Gauss' formula, this time resting on Helmholtz's theorem for
vector fields. Such a derivation, in turn, serves to shed light into the
connection existing between a conservation law like charge conservation and
the Maxwell equations. The key role played by the constitutive equations in
the construction of Maxwell's electromagnetism is briefly discussed, as well.
\end{abstract}

\section{Introduction}

\begin{figure}[ptb]
\begin{center}
\includegraphics[angle=0,scale=.6]{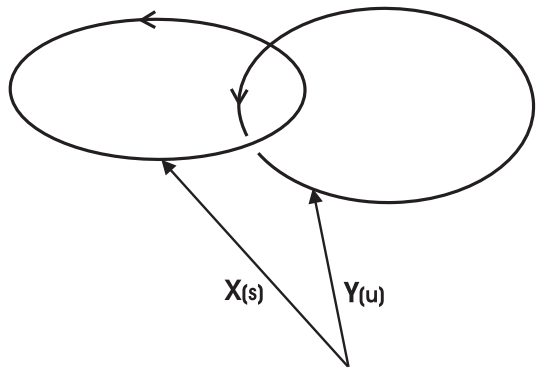}
%
%
%
%
%
%
%
%
%
%
%
%
\end{center}
\caption{Two non-intersecting curves $C_{1}:\mathbf{x}(s)$ and $C_{2}:%
\mathbf{y}(u)$, whose linking number is equal to one, in this case.}
\label{f1}
\end{figure}

Dated January 22, 1833, there is a note by Carl Friedrich Gauss that
appeared in Volume V of his complete works, a note that refers to a formula
which is nowadays known as the Gauss' linking integral. It represents the
very first result of what has grown to be in the following centuries a new
mathematical theory: knot theory, a branch of topology. Gauss presented his
result, a formula for calculating the linking number of two curves, without
giving any proof of it. In modern notation, what Gauss asserted was the
following. Consider two non-intersecting curves, $C_{1}:\mathbf{x}(s)$ and $%
C_{2}:\mathbf{y}(u)$, (see Fig.~\ref{f1}), and define the \emph{linking
integral} of the two curves, $Lk(C_{1},C_{2})$, to be given by

\begin{equation}
Lk(C_{1},C_{2})=\frac{1}{4\pi}\int_{C_{1}}\int_{C_{2}}\frac{\left( \mathbf{y}%
(u)-\mathbf{x}(s)\right) \cdot\left( \mathbf{y}^{\prime}(u)\times\mathbf{x}%
^{\prime}(s)\right) }{\left| \mathbf{y}(u)-\mathbf{x}(s)\right| ^{3}}duds,
\label{1}
\end{equation}
where the prime denotes derivation with respect to the curve parameter.

According to Gauss, when $C_{1}$ and $C_{2}$ are closed curves, $%
Lk(C_{1},C_{2})$ equals the number $n$ of times that one curve winds around
the other: $Lk(C_{1},C_{2})=n$ (amazingly, it is the factor $1/4\pi$ what
makes $Lk(C_{1},C_{2})$ an \emph{integer}-valued functional of the loops).
Furthermore, $Lk(C_{1},C_{2})$ is invariant under re-parametrization and
under continuous deformation of the curves.

In the course of time, Gauss' linking integral has been generalized within
the context of differential geometry\cite{flanders}. It has been defined for
any two oriented submanifolds in an Euclidean $n$-space $\mathbf{E}^{n}$,
whose respective dimensions, $r$ and $s$, have to satisfy $r+s+1=n$. The
special case of two curves in an Euclidean $3$-dimensional space corresponds
to $r=s=1$ and $n=3$. The generalization is based on a concept called the
\emph{degree of a mapping}. In order to introduce such a concept, one has to
rely on several tools from modern differential geometry, and it is based on
these tools that a formal proof of Gauss' assertion has been given. However,
a formal demonstration of this sort seems to be completely un-illuminating.
Surely, Gauss' steps leading to equation (\ref{1}) must have followed a very
different path. Unfortunately, it seems difficult to find in current
literature an elementary derivation of equation (\ref{1}). It should be
mentioned that the editor of Gauss' \emph{Works} conjectured that the method
used by Gauss in arriving at equation (\ref{1}) relied on calculating the
line integral of the magnetic field due to a steady current. Such a
conjecture seems to be substantiated by the fact that Gauss' linking
integral is presented in the volume of his \emph{Works} dedicated to
mathematical physics, in the section of electromagnetism. However, Gauss'
interest in arriving at a formula for the linking number seems to stem from
his duties as an astronomer: he wanted to know which regions of the
celestial sphere had to be observed over a period of time, in order to
register the passing of an asteroid or other astronomical objects \cite
{epple}. The Earth's orbit defines one curve, whereas the orbit of the
celestial object defines a second one. The linking integral of these two
curves can be shown to determine whether the celestial object can be
observed or not. Within this context, a magnetic field does not play any
role, of course, and a derivation of equation (\ref{1}) from magnetostatics
seems unlikely. Whether Gauss relied or not on magnetostatics for deriving
his result will remain an open question probably forever. Anyhow, it has
been recently published an illustrative and instructive article \cite
{hirshfeld} in which the connection to magnetostatics is presented as an
adequate starting point for deriving Gauss' assertion.

Let us see in detail how the connection between magnetostatics and equation (%
\ref{1}) arises. Consider the line integral $\oint\mathbf{B}(\mathbf{y}%
)\cdot d\mathbf{y}$ of a magnetic field around a closed curve, $C_{2}:%
\mathbf{y}(u)$. The magnetic field is assumed to be produced by a steady
current $I$ flowing through another loop, $C_{1}:\mathbf{x}(s)$. The field $%
\mathbf{B}$ can be obtained by using the well-known Biot-Savart formula \cite
{reitz},

\begin{equation}
\mathbf{B}(\mathbf{y})=\frac{\mu_{0}I}{4\pi}\int_{C_{1}}d\mathbf{x\times}%
\frac{\mathbf{y-x}}{\left| \mathbf{y-x}\right| ^{3}}.  \label{bs}
\end{equation}

\bigskip It leads to the following expression for the line integral to be
calculated:

\begin{align}
\oint_{C_{2}}\mathbf{B}(\mathbf{y})\cdot d\mathbf{y} &
=\oint_{C_{2}}\oint_{C_{1}}\frac{\mu_{0}I}{4\pi}\frac{d\mathbf{x\times}%
\left( \mathbf{y-x}\right) }{\left| \mathbf{y-x}\right| ^{3}}\cdot d\mathbf{y%
}  \notag \\
& =\frac{\mu_{0}I}{4\pi}\oint_{C_{2}}\oint_{C_{1}}\frac{\left( \mathbf{y-x}%
\right) \cdot\left( d\mathbf{y\times}d\mathbf{x}\right) }{\left| \mathbf{y-x}%
\right| ^{3}},  \label{2}
\end{align}
where the last step follows by applying the cyclic property of the triple
product.

Alternatively, we can calculate $\oint_{C_{2}}\mathbf{B}(\mathbf{y})\cdot d%
\mathbf{y}$ by using Ampere's law:

\begin{equation}
\oint_{C_{2}}\mathbf{B}(\mathbf{y})\cdot d\mathbf{y=}\int_{S}\mathbf{\nabla
\times B\cdot dS}=\mu_{0}\int_{S}\mathbf{J\cdot dS}=\mu_{0}I_{encl}=\mu
_{0}nI.  \label{2a}
\end{equation}
Here, $I_{encl}$ means the total amount of current crossing the surface $S$.
Comparing equations (\ref{2}) and (\ref{2a}) we obtain Gauss' result,
equation (\ref{1}) with $Lk(C_{1},C_{2})=n$. We stress, however, that
equation (\ref{1}) involves nothing but geometrical (or topological)
properties of the curves, whereas equation (\ref{2}) and Ampere's law do
express some intrinsic property of the magnetic field.

What lies at the origin of \ the laws of Biot-Savart and Ampere are
Maxwell's equations. These equations express a\emph{\ law of Nature}. In
principle, we should not take them as a starting point, in order to
demonstrate a purely \emph{geometrical} property of two curves. It is
certainly appealing to consider that given a current in a loop, it produces
a magnetic field which winds around the loop. But - let us stress it again -
this follows as a consequence of a law of Nature, which rules how a magnetic
field can be produced by the flow of electric charges. At most, the picture
of a magnetic field winding around a circuit could have inspired Gauss to
arrive at his assertion. At proving it, however, he must have followed quite
another path, in accordance with the requirements of utmost rigor that he
put on any mathematical result\footnote{%
As just two examples of this, let us remind Gauss' unpublished work on
non-Euclidean geometry, that he developed 30 years before Bolyai and
Lobachevsky, as well as his also unpublished work on special complex
functions. These works remained unpublished because they did not satisfy the
requirement of utmost rigor, that Gauss so fervently demanded from his and
other's work.}. To some physicists, questions of mathematical rigor might
appear superfluous or not worth to deal with, particularly in cases where
one can arrive straightforwardly at a formula, by starting from well
established natural laws. To others, however, it results obvious that a
mathematical truth cannot follow from a law of Nature. This, for the very
simple reason that what we call a natural law might change or become
obsolete in view of new facts, whereas a mathematical result, once it has
been proved, will remain true forever.

A parable might be useful to illustrate our concern. Let us assume that the
following series of events takes place:

1. In 1785 Coulomb reports his findings about a law of Nature, which now
bears his name: $F_{C}\sim q_{1}q_{2}/r^{2}$.

2. In 1833 Gauss reports a formula involving the sum of the angles of a
spherical triangle. However, he does not include a proof of his formula.

3. During the 20th century mathematicians develop differential geometry and
by applying this theory they prove Gauss' formula, establishing it as a
mathematical truth.

4. In 1998 a physics journal publishes a paper in which it is claimed that
Gauss' formula can be proved by simple means. Starting point of the proof is
Coulomb's law.

5. In 2020 very accurate experiments show beyond any reasonable doubt that
Coulomb's inverse-square law is not exact: It should be replaced by a power
law in which the exponent of $r$ is $-2.000000534(13)$.

We are obviously entitled to ask: Was the proof published in the physics
journal flawless? At first sight and in view of point $5$, it was not.

Now, let us put aside point $5$ - or assume it never happens - but ask anew:
\emph{Can} the proof published in the physics journal be correct? The mere
possibility that point $5$ \emph{could} happen makes the proof based on
Coulomb's law at least suspicious of being flawed. In any case, it raises
interesting questions: If the proof is correct, then it should be based only
apparently on Coulomb's law. Perhaps it was some geometrical property
lurking behind the inverse-square law what was actually used in the proof.
Or perhaps there exists some deep connection between Coulomb's law and
spherical symmetry that precludes point $5$ from ever occurring. Bringing
such a connection into light would certainly help saving time and efforts
dedicated to test the validity of Coulomb's law. For instance, sophisticated
experiments are continually planned to test natural laws like the
gravitational inverse-square law \cite{hoyle}. To be sure, as physicists we
are ready to accept the verdict of an experiment. But we must be able to
properly interpret what the verdict means and what it does not\footnote{%
As a matter of fact, a phenomenon like e.g. the Lamb shift can be
interpreted as an indirect demonstration that Coulomb's law does not hold
true at very small distances. Alternatively, it can be interpreted as
showing the limit of validity of a theory based on nonquantized fields.}.

One could still argue that we should distinguish between a physical law and
the mathematical statement in terms of which we express that physical law.
The physical law comes in when we assert that the real world behaves as
described by the mathematical statement. From this mathematical statement we
may deduce other such statements by a series of \emph{\ if-then}
relationships. In the case of Coulomb's law, such a relationship would tell
that \emph{if} the field of a point charge goes as $1/r^{2}$ \emph{then}
some other results follow. If some experiment then shows that the real world
does not behave according to the $1/r^{2}$ law, it does not mean that the
\emph{if-then} relationships turn to be false. This is true, of course.
However, without going into philosophical questions concerning the
foundations of mathematics, we may see it as embracing different theories
that can be constructed independently from one another, as free creations of
the human mind. Each mathematical theory consists of a set of axioms,
definitions, and theorems. It is clear that we cannot use a statement \emph{A%
} of one theory - be it an axiom or a theorem - to prove a statement \emph{B}
of another theory, unless \emph{A} is shared by both theories. Let one of
these theories be the mathematical idealization of the real world that we
may simply call, for the argument's sake, electrodynamics. Let the other
theory be pure geometry. We could hardly say that the $1/r^{2}$ statement of
Coulomb's law, or more generally, that Maxwell's equations fit into the body
of pure geometry. They are neither axioms nor theorems of pure geometry and
therefore cannot be used as starting points to prove purely geometrical
statements, like those involving the angles of a spherical triangle or the
properties of two curves.

Let us write Coulomb's law in vectorial form, $\mathbf{F}_{C}\sim q_{1}q_{2}(%
\mathbf{r}_{1}-\mathbf{r}_{2})/\left| \mathbf{r}_{1}-\mathbf{r}_{2}\right|
^{3}$, and so we have the integrand of Biot-Savart's law showing up. To put
Coulomb's law to a test is as justified as to put Biot-Savart's law to a
test. Coulomb's law can be traced back to one of Maxwell's equations: $%
\mathbf{\nabla \cdot E}=\rho /\varepsilon _{0}$. Biot-Savart's law can be
traced back to $\mathbf{\nabla }\times \mathbf{B=}\mu _{0}\mathbf{J}$, a
special, steady-state, case of one of the Maxwell's equations. Change the
spherical triangle in the parable above by Gauss' linking integral and the
foregoing remarks apply to the real case. This real case offers us the
opportunity to analyze the content of Maxwell's equations from a somewhat
new perspective. In the following, we will undertake an analysis of the
Biot-Savart law and the Maxwell's equations, on which the proof of Gauss'
linking integral that appeared in the real physics journal \cite{hirshfeld}
was based. This will lead us, at the end, to find the connection between the
principle of charge conservation and Maxwell's equations.

\section{\protect\bigskip Analysis of the Biot-Savart law}

The argument which allegedly proves Gauss' result was based on equations (%
\ref{2}) and (\ref{2a}). Let us see the extent to which such a proof
effectively depends on electromagnetism. Assuming that the derivation is
correct, it should be possible to reformulate it in terms of purely
geometrical concepts. In order to do this, we first try to see the essence
of the arguments: equation (\ref{2}) derives from the expression that gives
the magnetic field at point $\mathbf{y}$ produced by a steady
current-density $\mathbf{J}(\mathbf{x})$:

\begin{equation}
\mathbf{B}(\mathbf{y})=\frac{\mu_{0}}{4\pi}\int_{V}\mathbf{J}(\mathbf{x}%
)\times\frac{\mathbf{y-x}}{\left| \mathbf{y-x}\right| ^{3}}d^{3}x.  \label{3}
\end{equation}

Indeed, by taking $\mathbf{J}$ as appropriate for the current in a loop, the
volume integral in equation (\ref{3}) reduces to a line integral ($\mathbf{J}%
(\mathbf{x})d^{3}x\rightarrow Id\mathbf{x}$):

\begin{equation}
\mathbf{B}(\mathbf{y})=\frac{\mu_{0}I}{4\pi}\int_{C}d\mathbf{x\times}\frac{(%
\mathbf{y-x})}{\left| \mathbf{y-x}\right| ^{3}},  \label{3a}
\end{equation}
which is equation (\ref{bs}), the expression commonly referred to as the
Biot-Savart law \cite{reitz}.

Equation (\ref{3}), in turn, follows as a solution of one of the Maxwell's
equations (for steady fields):

\begin{equation}
\mathbf{\nabla}\times\mathbf{B=}\mu_{0}\mathbf{J},  \label{m1}
\end{equation}
together with $\mathbf{\nabla\cdot B}=0$, although this last equation is not
essential for our present purposes\footnote{%
Note that equation (\ref{3}) automatically satisfies $\mathbf{\nabla\cdot B}%
=0$. We remark that equation (\ref{3}) is a special solution of (\ref{m1}),
whereby the general solution can be obtained by adding to (\ref{3}) a term $%
\mathbf{\nabla}U$, with $U$ an arbitrary scalar function. For the purpose of
demonstrating Gauss' result we could have used the general solution, because
$\oint_{C}\mathbf{\nabla}U\cdot d\mathbf{y}=0$, so that it is not essential
to invoke $\mathbf{\nabla\cdot B}=0$.}.

Let us recapitulate the essential steps leading to Gauss' result. We have as
a starting point Maxwell's equation for a stationary current, equation (\ref
{m1}). From it, we derive Ampere's law, $\oint_{C_{2}}\mathbf{B}(\mathbf{y}%
)\cdot d\mathbf{y=}\mu_{0}\int_{S}\mathbf{J\cdot dS}$, as well as the
Biot-Savart law, equation (\ref{3a}), which leads to equation (\ref{2}). In
Ampere's law we must put $\int_{S}\mathbf{J\cdot dS=}I_{encl}$. Here, $%
I_{encl}$ means the \emph{total} current going through the loop $C_{2}$ that
encloses the surface $S$. In Biot-Savart's law, the current $I$ refers
instead to the current \emph{in} the wire. When this wire winds $n$ times
around $C_{2}$, we have $I_{encl}=nI$ and Gauss' result follows.

It seems therefore that the essential points in the foregoing proof of
Gauss' result are: 1) Maxwell's equation (\ref{m1}), and 2) the concept of
current as a flow of charged particles. Our question is: Is it possible to
extract from these two points some purely geometrical properties on which to
base an alternative proof?

Let us analyze more closely point 2). After all, the current might be merely
playing the role of a ``counting device'', and so it might be an
exchangeable unit within the foregoing reasoning. Alas, such a counting
property should not be intimately tied to the \emph{physical} nature of the
current. The factor $I$ appearing in equation (\ref{3a}) corresponds to the
current flowing through a thin wire, the loop $C_{1}$ in our case. The
replacement $\mathbf{J}(\mathbf{x})d^{3}x\rightarrow Id\mathbf{x}$ leading
from equation (\ref{3}) to (\ref{3a}) seems to be crucial. It follows from
the way we relate our concepts of \emph{current-density} and \emph{current}.
Indeed, such a replacement comes from considering that the volume element $%
d^{3}x$ appearing in equation (\ref{3}) can be written as $d^{3}x=S_{0}dx$,
where $S_{0}$ is the cross-sectional area of the wire and $dx$ a
distance-element along it (the direction of $dx$ being given through a unit
vector $\widehat{\mathbf{n}}$, parallel to $\mathbf{J}$). Thus, $\mathbf{J}(%
\mathbf{x})d^{3}x=\mathbf{J}(\mathbf{x})S_{0}dx=\left| \mathbf{J}(\mathbf{x}%
)\right| S_{0}\widehat{\mathbf{n}}dx=Id\mathbf{x}$, whereby the last step
expresses the connection between current-density and current: $\left|
\mathbf{J}(\mathbf{x})\right| S_{0}=I$.

Now, when going from equation (\ref{3}) to (\ref{3a}), the current $I$
refers to the charge going through each piece $d\mathbf{x}$ of the wire. On
the other hand, when we use equation (\ref{2a}) we must replace the
right-hand side, $\int_{S}\mathbf{J\cdot dS}$, by the \emph{total} current $%
I_{encl}$ crossing the surface $S$ which is enclosed by $C_{2}$. When the
current-carrying loop crosses $C_{2}$ a number $n$ of times, we must put $%
I_{encl}\equiv \int_{S}\mathbf{J\cdot dS}=nI$ in Ampere's law. As we said
before, replacing the left-hand side of equation (\ref{2}) by $\mu _{0}nI$
and cancelling the common factor $\mu _{0}I$, we obtain Gauss' result, $%
Lk(C_{1},C_{2})=n$. Thus, both the equation $I_{encl}=nI$ and the
relationship between current-density and current, as expressed by $\mathbf{J}%
(\mathbf{x})d^{3}x=Id\mathbf{x}$, seem to be intimately tied to our \emph{%
physical} concept of current as being a flow of particles. However, the
cancelation of $I$ in the last step of the proof seems to indicate that the
current was not so essential. We will see in the following that we can in
fact dispose of the concept of a current, and that we can replace Maxwell's
equation (\ref{m1}) by another one which expresses a purely geometrical
property.

\section{Derivation of the Gauss Linking Integral}

Let us now start by considering the closed curve $C_{1}:\mathbf{x}(s)$. This
curve can be embedded in a family of curves, under quite general conditions.
To define a family of curves amounts the same as to define a vector field $%
\mathbf{v}(\mathbf{r})$. Indeed, the curves pertaining to the family of
curves can be seen as \emph{integral curves} of the vector field, i.e., as
curves whose tangent vectors coincide with $\mathbf{v}(\mathbf{r})$. They
fulfill the following equation:

\begin{equation}
\frac{d\mathbf{x}(s)}{ds}=\mathbf{v}(\mathbf{x}(s)).  \label{8}
\end{equation}

We will be finally interested in only one member of the family of curves,
namely on $C_{1}$. In case that all the curves of the family are closed
curves, as we will assume henceforth, we have\footnote{%
We recall that the coordinate-free definition of the divergence of a vector
field $\mathbf{F}(\mathbf{x})$ is given by $\mathbf{\nabla}\cdot\mathbf{F}(%
\mathbf{x})=\lim_{V\rightarrow0}\frac{1}{V}\oint_{S}\mathbf{F}\cdot\mathbf{n}%
da$. When all integral curves of $\mathbf{F}(\mathbf{x})$ are closed, the
surface integral in the definition of the divergence gives zero.}

\begin{equation}
\mathbf{\nabla}\cdot\mathbf{v(r)}=0.  \label{9}
\end{equation}

Let us remark in passing that equation (\ref{9}) holds true also for the
velocity field of an incompressible fluid of constant and uniform density $%
\rho$. In absence of sources and sinks, such a velocity field satisfies the
continuity equation

\begin{equation}
\frac{\partial\rho}{\partial t}+\mathbf{\nabla}\cdot\rho\mathbf{v}=\rho%
\mathbf{\nabla}\cdot\mathbf{v}=0.  \label{10}
\end{equation}

Although in the present case equation (\ref{9}) follows because the integral
curves of $\mathbf{v}$ are closed, its relationship with the continuity
equation will be important afterwards. Now, for $\mathbf{v}$ satisfying
equation (\ref{9}) and under appropriate conditions (see Helmholtz's theorem
below), there exists a field $\mathbf{W}$, whose curl is $\mathbf{v}$:

\begin{equation}
\mathbf{\nabla}\times\mathbf{W=v.}  \label{11}
\end{equation}

This last equation is analogous to equation (\ref{m1}). We have so
introduced the vector fields $\mathbf{v}$ and $\mathbf{W}$, playing the
roles of $\mathbf{J}$ and $\mu_{0}^{-1}\mathbf{B}$, respectively. From
equation (\ref{11}) we can derive the analogous to Ampere's and Biot-Savart'
laws: equation (\ref{11}) can be solved for $\mathbf{W}$ explicitly, by
writing $\mathbf{W=\nabla\times Z}$, with

\begin{equation}
\mathbf{Z(y)}=\frac{1}{4\pi}\int_{V}\frac{\mathbf{v}(\mathbf{r})}{\left|
\mathbf{y-r}\right| }d^{3}r.  \label{12}
\end{equation}

From this $\mathbf{Z}$, by taking its curl, we obtain a solution of equation
(\ref{11}) given by

\begin{equation}
\mathbf{W(y)}=\frac{1}{4\pi}\int_{V}\mathbf{v}(\mathbf{r})\times \frac{(%
\mathbf{y-r)}}{\left| \mathbf{y-r}\right| ^{3}}d^{3}r.  \label{13}
\end{equation}
It corresponds to equation (\ref{3}) for the magnetic field. In order to
derive Gauss' result from here, by following similar steps as before, we
would need to reduce the volume integral in equation (\ref{13}) to a line
integral. To this end, we invoke the Dirac delta function to single out the
contribution to $\mathbf{W(y)}$ which comes from a single integral curve $%
\mathbf{x(}s\mathbf{)}$ of $\mathbf{v}(\mathbf{r})$. This is equivalent to
restricting the field $\mathbf{v}(\mathbf{r})$ in equation (\ref{13}) to be
different from zero only along the curve $\mathbf{x(}s\mathbf{)}$\footnote{%
As in the commonly employed definition of a current density $\mathbf{j}(%
\mathbf{x},t)$ due to a single charge moving along the path $\mathbf{r}(t)$,
i.e., $\mathbf{j}(\mathbf{x},t)=e\mathbf{v}(t)\delta^{3}(\mathbf{x}-\mathbf{r%
}(t))$, an expression like equation (\ref{15}) makes sense only under the
assumption that subsequent integrations over its arguments have to be
carried out. Equation (\ref{15}) is similar to the covariant form of the $4$%
-vector current density $j^{\alpha}(x)$, given in terms of the path $r(\tau)$
that is traced back by a charge in space-time: $j^{\alpha}(x)=ec\int d\tau
v^{\alpha}(\tau)\delta ^{(4)}(x-r(\tau))$.}:

\begin{equation}
\mathbf{v}(\mathbf{r})=\int\frac{d\mathbf{x}(s)}{ds}\delta^{3}(\mathbf{r}-%
\mathbf{x(}s\mathbf{)})ds.  \label{15}
\end{equation}

Introducing this expression for $\mathbf{v}(\mathbf{r})$ into equation (\ref
{13}) we obtain

\begin{align}
\oint_{C_{2}}\mathbf{W}(\mathbf{y})\cdot d\mathbf{y} & =\frac{1}{4\pi}%
\oint_{C_{2}}\left( \int_{V}\oint_{C_{1}}ds\frac{d\mathbf{x}(s)}{ds}\times%
\frac{\mathbf{y}-\mathbf{r}}{\left| \mathbf{y}-\mathbf{r}\right| ^{3}}%
\delta^{3}(\mathbf{r}-\mathbf{x}(s))d^{3}r\right) \cdot d\mathbf{y}  \notag
\\
& =\frac{1}{4\pi}\oint_{C_{2}}\left( \oint_{C_{1}}ds\frac{d\mathbf{x}(s)}{ds}%
\times\frac{\mathbf{y}-\mathbf{x}(s)}{\left| \mathbf{y}-\mathbf{x}(s)\right|
^{3}}\right) \cdot d\mathbf{y}.  \label{16}
\end{align}

We have, therefore, that

\begin{equation}
\oint_{C_{2}}\mathbf{W}(\mathbf{y})\cdot d\mathbf{y}=\frac{1}{4\pi}%
\oint_{C_{2}}\left( \oint_{C_{1}}d\mathbf{x}\times\frac{\mathbf{y}-\mathbf{x}%
}{\left| \mathbf{y}-\mathbf{x}\right| ^{3}}\right) \cdot d\mathbf{y}.
\label{16b}
\end{equation}
This last expression can in turn be changed as follows, by applying the
cyclic property of the triple product:

\begin{align}
\oint_{C_{2}}\mathbf{W}(\mathbf{y})\cdot d\mathbf{y} & =\frac{1}{4\pi}%
\oint_{C_{1}}\oint_{C_{2}}\frac{\left( \mathbf{x}-\mathbf{y}\right) \mathbf{%
\cdot}\left( d\mathbf{x}\times d\mathbf{y}\right) }{\left| \mathbf{x}-%
\mathbf{y}\right| ^{3}}  \notag \\
& =\frac{1}{4\pi}\oint_{C_{1}}\oint_{C_{2}}\frac{\left( \mathbf{y}(u)-%
\mathbf{x}(s)\right) \cdot\left( \mathbf{y}^{\prime}(u)\times \mathbf{x}%
^{\prime}(s)\right) }{\left| \mathbf{y}(u)-\mathbf{x}(s)\right| ^{3}}duds.
\label{17}
\end{align}

We have so obtained the right-hand side of Gauss' formula. In this case we
have made no reference to a current. As a second step, let us turn our
attention to the left-hand side of equation (\ref{17}). We have, by applying
Stokes theorem and considering equation (\ref{11}), that

\begin{equation}
\oint_{C_{2}}\mathbf{W(y)}\cdot d\mathbf{y}=\int_{S}\mathbf{(\nabla \times W)%
}\cdot \widehat{\mathbf{n}}dS=\int_{S}\mathbf{v}\cdot \widehat{\mathbf{n}}dS,
\label{flux}
\end{equation}
$S$ being the area enclosed by $C_{2}$. We have to prove that the integral $%
\int_{S}\mathbf{v}\cdot \widehat{\mathbf{n}}dS$, with $\mathbf{v}$ given by
equation (\ref{15}), readily equals $n$, the number of times that one curve
winds around the other, i.e., we have to demonstrate $\int_{S}\mathbf{v}%
\cdot \widehat{\mathbf{n}}dS=n$. If we succeed, we have the aforementioned
``counting device'' at our disposal, and with it the second piece that is
needed for the proof. Because the proof that $\int_{S}\mathbf{v}\cdot
\widehat{\mathbf{n}}dS=n$ is somewhat involved, we relegate it to an
Appendix.

Alternatively, we may introduce a small change in what we have done so far,
in order to give a proof of Gauss' result that resembles closely the one
based on magnetostatics. However, it is based on the concept of the \emph{%
strength} of a tube of curves, a concept whose properties might be difficult
to substantiate by purely geometrical means. It makes possible to show that $%
\int_{S}\mathbf{v}\cdot \widehat{\mathbf{n}}dS$ works as a ``counting
device''. Indeed, the current-carrying wire may be replaced by a tube of
curves, something that depends on $\mathbf{v(r)}$ alone: To this end,
consider some closed curve $\mathcal{C}$ in the region where $\mathbf{v(r)}$
is defined. Consider further a family of integral curves of $\mathbf{v(r)}$
passing through $\mathcal{C}$. These curves constitute a \emph{tube} of
curves. Now, the flux of $\mathbf{v(r)}$ across the surface $S$ bounded by $%
\mathcal{C}$ remains constant along the tube. That is, $\int_{S}\mathbf{v}%
\cdot \widehat{\mathbf{n}}dS=\int_{S^{\prime }}\mathbf{v}\cdot \widehat{%
\mathbf{n}}^{\prime }dS^{\prime }$ for all surfaces $S^{\prime }$ that cut
the tube. This result follows from $\mathbf{\nabla }\cdot \mathbf{v}=0$
alone. Let us refer to the flux $\sigma \equiv \int_{S}\mathbf{v}\cdot
\widehat{\mathbf{n}}dS$ as the \emph{strength} of the tube \cite{batchelor}.
It is always possible - with an appropriate redefinition of $\mathbf{v}$, if
necessary - to let the tube of curves shrink into a line, keeping fixed the
value of $\sigma $. It is clear that $\sigma $ can play just the same role
that the current $I$ has played before. That is, we may substitute $\mathbf{%
v(r)}d^{3}r$ by $\sigma d\mathbf{x}$, with $d\mathbf{x}$ a length element of
the curve $\mathbf{x}(s)$, and then proceed further as we did before:
equation (\ref{13}) reduces in this case to

\begin{equation}
\mathbf{W(y)}=\frac{\sigma}{4\pi}\int_{C_{1}}d\mathbf{r}\times\frac {(%
\mathbf{y-r)}}{\left| \mathbf{y-r}\right| ^{3}},
\end{equation}
so that the right sides of equations (\ref{16b}) and (\ref{17}) acquire a
factor $\sigma$. If the curve $C_{1}:\mathbf{x}(s)$ crosses the surface $S$
enclosed by $C_{2}$ a number $n$ of times, it contributes with $n\sigma$ to
the total flux through $S$ that appears on the left side of equation (\ref
{17}). The common factor $\sigma$ cancels then, and we arrive at Gauss'
result again. However, this time the strength - at variance with the current
- does refer to a geometrical property alone.

By introducing the strength we are provided with a purely geometrical
concept on which our proof of Gauss' result can rely. Nevertheless, for the
sake of demonstrating Gauss' result, the introduction of the strength
appears still as a disposable artifice. As already said, it is in fact
possible to demonstrate that $\int_{S}\mathbf{v}\cdot\widehat{\mathbf{n}}%
dS=n $ without introducing the strength as a substitute of the current, as
it is done in the Appendix.

In any case, we have arrived at Gauss' result by applying purely geometrical
facts. We do not claim that the present method resembles more likely what
Gauss probably did, than the method conjectured by the editor of Gauss'
\emph{Works}. Although it is true that the genius of Gauss could have
anticipated some of Helmholtz's results needed for the proof given above,
there are other aspects that one should take into account before making any
conjecture about the line of reasoning actually followed by Gauss. We are
not concerned, however, with this kind of issues here.

Our analysis has brought into light that the proof of Gauss' result that was
based on Biot-Savart's and Ampere's laws was in fact based on Maxwell's
equation $\mathbf{\nabla}\times\left( \mu_{0}^{-1}\mathbf{B}\right) \mathbf{%
=J}$. We also saw that this last equation could be replaced, for the sake of
proving Gauss' result, by the equation $\mathbf{\nabla}\times \mathbf{W=v}$.
This equation can be seen as the one defining a vector field $\mathbf{W(r)}$%
, given another vector field $\mathbf{v(r)}$ whose divergence vanishes: $%
\mathbf{\nabla}\cdot\mathbf{v}=0$. This one restriction put on $\mathbf{v(r)}
$ is a particular instance of the continuity equation. We are so naturally
led to ask about the equation that would replace $\mathbf{\nabla }\times%
\mathbf{W=v}$, when instead of $\mathbf{\nabla}\cdot\mathbf{v}=0$ we require
from $\mathbf{v}$ to satisfy the continuity equation: $\partial _{t}\rho+%
\mathbf{\nabla}\cdot\rho\mathbf{v=}0$. Putting it otherwise, we are led to
ask about the connection between the continuity equation and Maxwell's
equations. This is the issue we want to discuss in what follows, with the
help of Helmholtz's theorem.

\section{Helmholtz's theorem}

In this Section we review the aforementioned Helmholtz's theorem \cite
{hauser,kobe1,kobe2,baierlein,arfken}. According to this theorem, a vector
field $\mathbf{v}(\mathbf{r})$ is completely determined by its divergence
and its curl, together with a boundary condition which specifies the normal
component of the field, $\widehat{\mathbf{n}}\cdot\mathbf{v}$, at the
boundary of the domain where the vector field is to be determined. For
physical applications it is natural to take as ``boundary'' the infinity and
the vector field vanishing there. Helmholtz's theorem then says that we can
write $\mathbf{v(r)}$ in terms of two potentials, $\varphi$ and $\mathbf{A}$%
, as

\begin{equation}
\mathbf{v(r)}=-\mathbf{\nabla}\varphi+\mathbf{\nabla}\times\mathbf{A,}
\label{34}
\end{equation}
where $\varphi$ and $\mathbf{A}$ can be expressed through the divergence and
the curl of $\mathbf{v}(\mathbf{r})$. Writing

\begin{align}
\mathbf{\nabla}\cdot\mathbf{v}(\mathbf{r}) & =\rho(\mathbf{r})  \label{35} \\
\mathbf{\nabla}\times\mathbf{v}(\mathbf{r}) & =\mathbf{j}(\mathbf{r})
\label{35a}
\end{align}
we have that

\begin{equation}
\mathbf{v}(\mathbf{r}_{1})=-\mathbf{\nabla}\int d^{3}r_{2}\frac{\rho (%
\mathbf{r}_{2})}{4\pi\left| \mathbf{r}_{1}-\mathbf{r}_{2}\right| }+\mathbf{%
\nabla}\times\int d^{3}r_{2}\frac{\mathbf{j}(\mathbf{r}_{2})}{4\pi\left|
\mathbf{r}_{1}-\mathbf{r}_{2}\right| }.  \label{36}
\end{equation}

This can be written, alternatively, as

\begin{equation}
\mathbf{v}(\mathbf{r}_{1})=\mathbf{\nabla}\int d^{3}r_{2}G(\mathbf{r}_{1},%
\mathbf{r}_{2})\rho(\mathbf{r}_{2})-\mathbf{\nabla}\times\int d^{3}r_{2}G(%
\mathbf{r}_{1},\mathbf{r}_{2})\mathbf{j}(\mathbf{r}_{2}),  \label{37}
\end{equation}
with the Green's function $G(\mathbf{r}_{1},\mathbf{r}_{2})$ satisfying

\begin{equation}
\nabla^{2}G(\mathbf{r}_{1},\mathbf{r}_{2})=\delta(\mathbf{r}_{1}-\mathbf{r}%
_{2})  \label{38}
\end{equation}
and vanishing at infinity. Under this last condition, the solution of
equation (\ref{38}) is given explicitly by

\begin{equation}
G(\mathbf{r}_{1},\mathbf{r}_{2})=-\frac{1}{4\pi\left| \mathbf{r}_{1}-\mathbf{%
r}_{2}\right| }.  \label{39}
\end{equation}

Now, assume that we prescribe only the divergence $\rho$ of a field, which
is a function not only of position but of time as well, i.e., $\rho =\rho(t,%
\mathbf{r})$. Let our boundary condition be that $\rho$ vanish at infinity.
Helmholtz's theorem states that there is a field, call it $\mathbf{D}$, such
that

\begin{equation}
\mathbf{\nabla}\cdot\mathbf{D}(t,\mathbf{r})=\rho(t,\mathbf{r}).  \label{39a}
\end{equation}
\

The field $\mathbf{D}(t,\mathbf{r})$ is explicitly given by

\begin{equation}
\mathbf{D}(t,\mathbf{r})=-\mathbf{\nabla}_{\mathbf{r}}\int d^{3}r_{1}\frac{%
\rho(t,\mathbf{r}_{1})}{4\pi\left| \mathbf{r}-\mathbf{r}_{1}\right| }+%
\mathbf{\nabla}_{\mathbf{r}}\times\mathbf{Z}_{1},  \label{40}
\end{equation}
with $\mathbf{Z}_{1}$ an arbitrary field that we may take equal to zero, for
simplicity. Note that the time $t$ plays, in all of this, \emph{only the
role of a parameter} that can be appended to the fields, without having yet
any dynamical meaning. The field $\mathbf{D}(t,\mathbf{r})$ satisfies
therefore the only condition we have put upon it, i.e., $\mathbf{\nabla}%
\cdot \mathbf{D}(t,\mathbf{r})=\rho(t,\mathbf{r}).$ Its curl has been
assumed as unspecified or else set equal to zero. We have then, by applying
the gradient operator that appears in equation (\ref{40}),

\begin{equation}
\mathbf{D}(t,\mathbf{r})=\frac1{4\pi}\int d^{3}r_{1}\frac{\rho(t,\mathbf{r}%
_{1})}{\left| \mathbf{r}-\mathbf{r}_{1}\right| ^{2}}\frac{\left( \mathbf{r}-%
\mathbf{r}_{1}\right) }{\left| \mathbf{r}-\mathbf{r}_{1}\right| }.
\label{41}
\end{equation}

For a point-like charge we set $\rho(t,\mathbf{r})=q\delta(\mathbf{r}-%
\mathbf{r}_{0}(t))$ and the above expression reduces to

\begin{equation}
\mathbf{D}(t,\mathbf{r})=\frac{q}{4\pi\left| \mathbf{r}-\mathbf{r}%
_{0}(t)\right| ^{2}}\frac{\mathbf{r}-\mathbf{r}_{0}(t)}{\left| \mathbf{r}-%
\mathbf{r}_{0}(t)\right| }.  \label{41a}
\end{equation}

According to equations (\ref{41}) and (\ref{41a}), the field $\mathbf{D}(t,%
\mathbf{r})$ at time $t$ corresponds to an instantaneous Coulomb field
produced by a charge distribution $\rho$, or else by a point-like charge $q$%
. Such a result would correspond to an instantaneous response of the field
to any change suffered by the charge distribution. Apparently, there is here
a contradiction with the finite propagation-time needed by any signal. This
issue has been discussed and cleared, in the case of the \emph{complete} set
of Maxwell's equations, by showing that both the longitudinal and the
transverse parts of the electric field contain such instantaneous
contributions, which turn out to cancel each other \cite{donnely}. Note that
by taking $\mathbf{Z}_{1}$ equal to zero in equation (\ref{40}) we have $%
\mathbf{\nabla}\times\mathbf{D=0}$ in our case, which is not what happens
when $\mathbf{D}$ has to satisfy (together with $\mathbf{H}$) the complete
system of Maxwell's equations.

Coming back to Helmholtz's theorem, assume next that $\rho(t,\mathbf{r})$
satisfies the continuity equation

\begin{equation}
\partial_{t}\rho+\mathbf{\nabla}\cdot\mathbf{j}=0.
\end{equation}

By using equation (\ref{39a}) the continuity equation can be written as

\begin{equation}
\mathbf{\nabla}\cdot\left( \partial_{t}\mathbf{D}+\mathbf{j}\right) =0.
\label{42a}
\end{equation}

Here, again, we can apply Helmholtz's theorem and write the field $%
\partial_{t}\mathbf{D}+\mathbf{j}$ as the curl of some other field $\mathbf{H%
}$, whose divergence we do not need to specify. This field is given by
Helmholtz's theorem as

\begin{equation}
\mathbf{H}(t,\mathbf{r})=\mathbf{\nabla}_{\mathbf{r}}\times\int d^{3}r_{2}%
\frac{(\mathbf{j(}t,\mathbf{r}_{2}\mathbf{)}+\partial_{t}\mathbf{D(}t,%
\mathbf{r}_{2}\mathbf{)})}{4\pi\left| \mathbf{r}-\mathbf{r}_{2}\right| }+%
\mathbf{\nabla}\varphi.  \label{43}
\end{equation}

As before, we may take $\mathbf{\nabla}\cdot\mathbf{H}=0$, and hence $%
\varphi=0\mathbf{.}$ That such $\mathbf{H}(t,\mathbf{r})$, with $\varphi$
set equal to zero, fulfills the required equation, namely

\begin{equation}
\mathbf{\nabla}\times\mathbf{H}=\mathbf{j}+\partial_{t}\mathbf{D},
\label{44}
\end{equation}
can be shown as follows. By writing

\begin{equation}
\mathbf{A}(t,\mathbf{r})\equiv\int d^{3}r_{2}\frac{(\mathbf{j(}t,\mathbf{r}%
_{2}\mathbf{)}+\partial_{t}\mathbf{D(}t,\mathbf{r}_{2}\mathbf{)})}{%
4\pi\left| \mathbf{r}-\mathbf{r}_{2}\right| }  \label{45}
\end{equation}
in equation (\ref{43}), we have $\mathbf{H}(t,\mathbf{r})=\mathbf{\nabla }_{%
\mathbf{r}}\times\mathbf{A}(t,\mathbf{r})$, so that taking the curl on both
sides of this last equation we obtain

\begin{equation}
\mathbf{\nabla}_{\mathbf{r}}\times\mathbf{H(}t\mathbf{,r)=\nabla}_{\mathbf{r}%
}\times\mathbf{\nabla}_{\mathbf{r}}\times\mathbf{A(}t\mathbf{,r)=\nabla
_{r}(\nabla_{r}\cdot A)-\nabla}_{\mathbf{r}}^{2}\mathbf{A.}  \label{46}
\end{equation}

The first term on the right-hand side vanishes:

\begin{equation}
\mathbf{\nabla}_{\mathbf{r}}\mathbf{(\nabla_{r}\cdot A)}=\mathbf{0.}
\label{47}
\end{equation}

This can be seen by applying the operator $\mathbf{\nabla _{r}(\nabla
_{r}\cdot )}$ to $\mathbf{A}$ given by equation (\ref{45}). One replaces the
second derivatives with respect to $\mathbf{r}$ in the term $\left| \mathbf{r%
}-\mathbf{r}_{2}\right| ^{-1}$ inside the integral, by second derivatives
with respect to $\mathbf{r}_{2}$. Integrating by parts one obtains equation (%
\ref{47}), upon using that the field $\mathbf{j(}t,\mathbf{r)}+\partial _{t}%
\mathbf{D(}t,\mathbf{r)}$ is solenoidal, i.e., one whose divergence
vanishes, and that it is bounded in space, or that it vanishes faster than $%
r^{-1}$ for large $r$ \cite{arfken}. We have therefore $\mathbf{\nabla }%
\times \mathbf{H=-\nabla }^{2}\mathbf{A}$, so that

\begin{equation}
\mathbf{\nabla}_{\mathbf{r}}\times\mathbf{H}=\mathbf{-}\int d^{3}r_{2}(%
\mathbf{j(}t,\mathbf{r}_{2}\mathbf{)}+\partial_{t}\mathbf{D(}t,\mathbf{r}_{2}%
\mathbf{)})\mathbf{\nabla}_{\mathbf{r}}^{2}\left( \frac {1}{4\pi\left|
\mathbf{r}-\mathbf{r}_{2}\right| }\right) .  \label{47a}
\end{equation}

Taking now into account that
\begin{equation}
\mathbf{\nabla}_{\mathbf{r}}^{2}\left( \frac{1}{\left| \mathbf{r}-\mathbf{r}%
_{2}\right| }\right) =-4\pi\delta\left( \mathbf{r}-\mathbf{r}_{2}\right) ,
\label{47b}
\end{equation}
we finally obtain

\begin{equation}
\mathbf{\nabla}\times\mathbf{H}=\mathbf{j(}t,\mathbf{r)}+\partial _{t}%
\mathbf{D(}t,\mathbf{r),}  \label{47c}
\end{equation}
which is identical to one of the two Maxwell's equations with sources. The
other Maxwell equation that has a source term is identical to equation (\ref
{39a}).

\section{Maxwell's equations with sources}

Let us summarize how the inhomogeneous Maxwell's equations followed from the
assumption of charge conservation. Let us describe charge - or any other
quantity, e.g., matter - by its density $\rho (t,\mathbf{x})$ and its
velocity distribution $\mathbf{v}(t,\mathbf{x})$. It is well known that
charge or matter conservation can be expressed through the equation of
continuity

\begin{equation}
\partial_{t}\rho+\mathbf{\nabla}\cdot(\rho\mathbf{v})=0.  \label{17a}
\end{equation}

As we have shown above, given a bounded, scalar, function $\rho(t,\mathbf{x}%
) $, there exists a field $\mathbf{Y}(t,\mathbf{x})$, satisfying

\begin{equation}
\mathbf{\nabla}\cdot\mathbf{Y}=\rho,  \label{18}
\end{equation}
so that equation (\ref{17a}) can be written in the form

\begin{equation}
\mathbf{\nabla}\cdot(\partial_{t}\mathbf{Y}+\rho\mathbf{v})=0.  \label{19}
\end{equation}

Because the field $\partial_{t}\mathbf{Y}+\rho\mathbf{v}$ is divergenceless,
there exists another field $\mathbf{W(}t,\mathbf{x)}$ such that

\begin{equation}
\mathbf{\nabla}\times\mathbf{W=}\partial_{t}\mathbf{Y}+\rho\mathbf{v.}
\label{20}
\end{equation}

Equations (\ref{18}) and (\ref{20}), written with the usual notation ($%
\mathbf{Y\rightarrow D}$, $\mathbf{W\rightarrow H}$, $\rho \mathbf{%
v\rightarrow j}$), i.e.,

\begin{align}
\mathbf{\nabla}\cdot\mathbf{D} & =\rho  \label{21} \\
\mathbf{\nabla}\times\mathbf{H} & =\mathbf{j}+\partial_{t}\mathbf{D,}
\label{21a}
\end{align}
are the two inhomogeneous Maxwell's equations. They follow from equation (%
\ref{17a}) alone. Reciprocally, the continuity equation follows from
equations (\ref{21}) and (\ref{21a}) by taking the divergence of the second
of these equations and replacing in it the first equation. We stress that
the curl of $\mathbf{D}$ and the divergence of $\mathbf{H}$ have been left
unspecified - or set arbitrarily equal to zero - when going from equation (%
\ref{17a}) to equations (\ref{21}) and (\ref{21a}). In the case of the
electromagnetic field, such an arbitrariness does not correspond to the
observed facts, as charge and current distributions completely determine the
fields they produce. There must be, therefore, additional equations for the
curl of $\mathbf{D}$ and the divergence of $\mathbf{H}$ in order to
completely determine these fields, when their boundary conditions are also
specified.

Let us now turn our attention to the source-free, or homogeneous, Maxwell's
equations:

\begin{align}
\mathbf{\nabla}\cdot\mathbf{B} & =0  \label{22} \\
\mathbf{\nabla}\times\mathbf{E+}\partial_{t}\mathbf{B} & =\mathbf{0}.
\label{22a}
\end{align}

Of course, by a similar reasoning as above, we could derive these equations
from the conservation of something, call it $\rho_{m}$. The conservation of $%
\rho_{m}$ could be expressed through an equation similar to the equation of
continuity, equation (\ref{17a}). After deriving equations similar to (\ref
{21}) and (\ref{21a}) we put $\rho_{m}=0$ and obtain equations (\ref{22})
and (\ref{22a}). In other words, this something - a ``magnetic charge'' if
you want - is assumed to be conserved: zero before and zero after.

Alternatively, the source-free Maxwell's equations can be understood as a
purely mathematical statement telling us that the fields $\mathbf{E}$ and $%
\mathbf{B}$ derive from a scalar function $\varphi$ and a vector potential $%
\mathbf{A}$ (not to be confused with those in equation (\ref{34})), through
the definitions

\begin{align}
\mathbf{E} & =-\mathbf{\nabla}\varphi-\partial_{t}\mathbf{A,}  \label{23} \\
\mathbf{B} & =\mathbf{\nabla}\times\mathbf{A.}  \label{23a}
\end{align}

In other words, given two \emph{arbitrary} fields, $\varphi$ and $\mathbf{A}
$, by defining $\mathbf{E}$ and $\mathbf{B}$ through equations (\ref{23})
and (\ref{23a}), it follows that equations (\ref{22}) and (\ref{22a}) are
identically satisfied.

In any case, we should stress that equations (\ref{21}, \ref{21a}) on the
one hand, and equations (\ref{22}, \ref{22a}) on the other hand, are - up to
this point - independent from each other. We may connect them through some
\emph{constitutive equations}, like, e.g.,

\begin{align}
\mathbf{D} & =\varepsilon\mathbf{E}  \label{24} \\
\mathbf{H} & =\mu^{-1}\mathbf{B.}  \label{24a}
\end{align}

These equations are usually assumed to describe a linear medium of
electrical permittivity $\varepsilon $ and magnetic permeability $\mu $. A
particular case of such a medium is vacuum, and the system of equations (\ref
{21}, \ref{21a}, \ref{22}, \ref{22a}), that arises out of a connection like
the one given by equations (\ref{24}, \ref{24a}) is what we know as the
complete system of Maxwell's equations.

Let us stress once again that without connecting $\left( \mathbf{D}\text{, }%
\mathbf{H}\right) $ with $\left( \mathbf{E}\text{, }\mathbf{B}\right) $
through some constitutive equations we have no closed system. The equations
that we have written down for $\left( \mathbf{D}\text{, }\mathbf{H}\right) $%
, that is Maxwell's equations with sources, can also be written down for a
fluid, for example. These equations are, as already stated, $\mathbf{\nabla }%
\cdot\mathbf{Y}=\rho$ and $\mathbf{\nabla}\times\mathbf{W=j+}\partial _{t}%
\mathbf{Y}$. We can expect that any conclusion that can be derived in the
realm of electrodynamics from the equations $\mathbf{\nabla}\cdot \mathbf{D}%
=\rho$ and $\mathbf{\nabla}\times\mathbf{H}=\mathbf{j}+\partial _{t}\mathbf{D%
}$ without coupling them to the source-free Maxwell's equations, will have a
corresponding result in the realm of fluid dynamics. This assertion can be
illustrated by two examples: 1) The case in which a point-like singularity
of $\rho$ is produced in the interior of a fluid, for which one obtains a
velocity-field obeying a law that is mathematically identical to Coulomb's
law \cite{batchelor}. 2) The case of a fluid where a so-called vortex tube
appears (tornadoes and whirl-pools are associated phenomena), in which case
- after approximating the vortex-tube by a line singularity - one obtains a
velocity-field through an expression which is mathematically identical to
the Biot-Savart law \cite{batchelor}.

Finally, let us mention that the derivation of the inhomogeneous Maxwell's
equations as a consequence of charge conservation is nothing new \cite
{parrot}. It follows as a direct application of a theorem of de Rahm for
differential forms: given a $4$-vector $j^{\alpha}(x)$ for which a
continuity equation holds, $\partial_{\alpha}j^{\alpha}=0$, there exists an
antisymmetric tensor $F^{\alpha\beta}=-F^{\beta\alpha}$ such that $%
\partial_{\alpha }F^{\alpha\beta}=j^{\beta}$. This last equation is nothing
but the tensorial form of the inhomogeneous Maxwell's equations (\ref{21})
and (\ref{21a}). Now, the tensor $F^{\alpha\beta}$ does not need to be
derivable from a $4$-vector $A^{\alpha}$, according to $F^{\alpha\beta}=%
\partial^{\alpha}A^{\beta }-\partial^{\beta}A^{\alpha}$. For this to be the
case, $F^{\alpha\beta}$ would have to satisfy $\partial^{\alpha}F^{\beta%
\gamma}+\partial^{\beta }F^{\gamma\alpha}+\partial^{\gamma}F^{\alpha\beta}=0$%
. This is the tensorial form of the homogeneous Maxwell's equations (\ref{22}%
) and (\ref{22a}). In other words, given $j^{\alpha}$ and $A^{\alpha}$, with
$j^{\alpha}$ satisfying a continuity equation, we can introduce two
antisymmetric tensors, $F_{(1)}^{\alpha\beta}$ and $F_{(2)}^{\alpha\beta}$,
the first one satisfying $\partial_{\alpha}F_{(1)}^{\alpha\beta}=j^{\beta}$
and the second one, $F_{(2)}^{\alpha\beta}\equiv\partial^{\alpha}A^{\beta}-%
\partial^{\beta }A^{\alpha}$, satisfying $\partial^{\alpha}F_{(2)}^{\beta%
\gamma}+\partial^{\beta}F_{(2)}^{\gamma\alpha}+\partial^{\gamma}F_{(2)}^{%
\alpha\beta }=0$. In order that these two equations do conform a closed
system, i.e. the\emph{\ total }system of Maxwell's equations, we need to
connect $F_{(1)}^{\alpha\beta}$ with $F_{(2)}^{\alpha\beta}$ through some
constitutive relation\footnote{%
As a matter of fact, it is not necessary to rest on de Rham's theorem and
the theory of differential forms on manifolds in order to derive the
foregoing conclusions in tensorial form. Indeed, one can start with the
tensorial form of Helmholtz's theorem (see Ref.\cite{kobe1}) and go-ahead
with a similar reasoning as the one we have followed in the preceding
paragraphs.}.

\section{Summary and Conclusions}

We have presented an elementary proof of Gauss' result concerning the
linking number of two curves. The proof is based on geometrical properties
of a vector field, properties that are at the root of Helmholtz's theorem.
It could be cleared up what at first glance appeared as an astonishing
connection between Gauss' linking integral of two curves and the line
integral of a magnetic field. By exposing the connection that holds between
Helmholtz's theorem and the linking integral of Gauss, we could show which
purely geometrical properties were behind the line integral of a magnetic
field. In this way, we could lay bare the essential constituents of the
connection between an empirically established natural law (Maxwell's
equations) and a geometrical property of two curves (their linking
integral). Such essential constituents are certainly not characteristic of
the electromagnetic field. As a consequence, we showed that starting from
the conservation of any property - like mass or charge - for which the
continuity equation holds true, one can arrive at equations which are
mathematically identical to the Maxwell's equations. These equations are
therefore tightly linked to a general statement telling us that something is
conserved. Consider anything - charge, matter, or whatever - the amount of
which that is contained inside an arbitrary volume changes with time. If
this change is exclusively due to a flow through the volume's boundary, then
a continuity equation holds true. As a consequence of it, a pair of
Maxwell-like equations must be fulfilled for some auxiliary fields that we
may introduce, playing the role of $\mathbf{D}$ and $\mathbf{H}$.

The fact that Maxwell's equations with sources can be derived from charge
conservation is known \cite{parrot}, although it is not usual to find it
mentioned in standard textbooks of electromagnetism. Maxwell's equations
with sources involve the fields $\mathbf{D}$ and $\mathbf{H}$, whereas the
source-free equations involve the fields $\mathbf{E}$ and $\mathbf{B}$. It
is through some constitutive equations connecting $\left( \mathbf{D}\text{, }%
\mathbf{H}\right) $ with $\left( \mathbf{E}\text{, }\mathbf{B}\right) $ that
we obtain a closed system, i.e., the complete system of Maxwell's equations.
The constitutive equations express, in some way, the underlying properties
of the medium where the fields act or are produced. From this perspective,
Maxwell's equations entail besides charge conservation some other properties
of the medium, yet to be unraveled. These properties are effectively
described, in the simplest case, through the permittivity $\varepsilon $ and
the permeability $\mu $. The first one refers to electrical, the second one
to magnetic, properties of the medium, be it vacuum or any other one. It is
just when the equations for $\left( \mathbf{D}\text{, }\mathbf{H}\right) $
together with those for $\left( \mathbf{E}\text{, }\mathbf{B}\right) $ do
form a closed system, that we can derive a wave equation for these fields.
The velocity of wave propagation is then given by $c=(\varepsilon \mu
)^{-1/2}$, the velocity of light. It is remarkable that the velocity of
light can be decomposed in terms of a product of two independent parameters.
However, the historical development in physics has led us to look at $c$ as
a fundamental constant of Nature, instead of $\varepsilon $ and $\mu $ .
Nevertheless, currently discussed and open questions related to accelerated
observers, Unruh radiation, self-force on a charge, magnetic monopoles and
the like, might well require an approach where the role of $c$ recedes in
favor of quantities like $\varepsilon $ and $\mu $. Such an approach could
perhaps be reminiscent of Helmholtz's efforts to develop an ether theory as
the basis of electrodynamics, a theory that strongly relied on the
mathematical machinery that Gauss so much helped to develop. If this happens
to be the case, it will seem somewhat ironical that the Maxwell's equations,
when written in the - by now - most commonly used \emph{Gaussian} units, do
not include but the single constant $c$, hiding so $\varepsilon $ and $\mu $
from our view. These last two constants might well be key pieces that remain
buried under the beauty of a unified theory of electromagnetic phenomena,
which is the version of electrodynamics that we know and use today.

\section{Appendix}

The assertion that $\int_{S}\mathbf{v}\cdot\widehat{\mathbf{n}}dS=n$ can be
proved in the following way, in which we will try not to veil with abstract
definitions and too much mathematical detail what we do intuitively take for
true.

The curve $C_{1}$ pierces the surface $S$ at different points $P_{1},$ $%
P_{2},\ldots,P_{N}$. These points can be obtained as intersections of $C_{1}$
and $S$\footnote{%
Let $C_{1}$ be given by equations of the form $x^{i}=f^{i}(s)$, and $S$ by
equations of the form $x^{i}=g^{i}(u,v)$ with $i=1,2,3$ and $s,u,v$ some
parameters. The points of intersection $P_{k}\in$ $C_{1}\cap S$ are obtained
by solving the system $f^{i}(s)=g^{i}(u,v)$, $i=1,2,3,$ for $s,$ $u,$ $v$.
In general, there are a number $N\geq0$ of solutions $\left(
s_{k},u_{k},v_{k}\right) $, $k=1,\ldots N,$ of these equations.}. Consider a
neighborhood $\mathcal{N}_{i}$ of $P_{i}$, it being sufficiently small so as
to contain no other point $P_{j}\neq P_{i}.$ We can deform this small
portion $\mathcal{N}_{i}$ of $S$, making it flat and lying parallel to the $%
XY$-plane. By so doing we do not change the value of $\int_{S}\mathbf{v}%
\cdot \mathbf{n}dS$, as long as $C_{2}$, the bounding curve of $S$, remains
unchanged. The idea is to split the integral $\int_{S}\mathbf{v}\cdot\mathbf{%
n}dS$ into $N$ contributions, each one of them corresponding to a small,
flat surface that locally can be made lye parallel to the $XY$-plane. Now,
in general, the integral $\int_{S}\mathbf{v}\cdot\mathbf{n}dS$ can be
expressed as \cite{nikolsky}

\begin{equation}
\int_{S}\mathbf{v}\cdot\mathbf{n}dS=\int_{S}v_{z}(x,y,f_{3}(x,y))dxdy\text{.}
\tag{A.1}  \label{cartes1}
\end{equation}
Here, $\mathbf{v}=(v_{x},v_{y},v_{z})$, whereas $S$ is assumed to be
described by the equation $z=f_{3}(x,y)$. With $\mathbf{v}$ given by
equation (\ref{15}) and after properly adapting the variables in equation (%
\ref{cartes1}) to our case, we have

\begin{equation}
\int_{S}\mathbf{v}\cdot\mathbf{n}dS=\int_{S}\int ds\frac{dz}{ds}\delta
(r_{x}-x(s))\delta(r_{y}-y(s))\delta(r_{z}-z(s))dr_{x}dr_{y}.  \tag{A.2}
\label{cartes2}
\end{equation}

In order to evaluate the integral on the right-hand side of\ equation (\ref
{cartes2}) we use the rule

\begin{equation}
\delta(g(x))=\sum_{x_{i}}\frac{\delta(x-x_{i})}{\left|
g^{\prime}(x_{i})\right| },  \tag{A.3}  \label{delta}
\end{equation}
where the $x_{i}$ satisfy $g(x_{i})=0$. Applying this property successively
to equation (\ref{cartes2}) we obtain

\begin{align*}
\int_{S}\mathbf{v}\cdot \mathbf{n}dS& =\sum_{i=1}^{N}\int_{\mathcal{N}%
_{i}}\int ds\frac{dz}{ds}\delta (r_{x}-x(s))\delta (r_{y}-y(s))\delta \left(
r_{z}(r_{x},r_{y})-z(s)\right) dr_{x}dr_{y} \\
& =\sum_{i=1}^{N}\int_{\mathcal{N}_{i}}\frac{dz/ds}{\left| dz/ds\right| }%
\delta (r_{x}-x(s))\delta (r_{y}-y(s))dr_{x}dr_{y} \\
& =\sum_{i=1}^{N}\pm \int_{\mathcal{N}_{i}}\delta (r_{x}-x(s))\delta
(r_{y}-y(s))dr_{x}dr_{y}=\sum_{i=1}^{N}\pm 1=n\leq N.
\end{align*}

We have used the fact that the surface integral splits into $N$ integrals,
i.e., as many as the number of times that $C_{1}$ pierces $S$. We have also
used that each $\mathcal{N}_{i}$ can be described as a small part of the
plane $r_{z}=const.$. It might occur that in some cases $C_{1}$ goes through
$S$ in one direction, turns then back and goes through $S$ again but in the
opposite direction, without having wound around the boundary $C_{2}$ of $S$
in-between. For these cases we have a positive contribution ($dz/ds=+\left|
dz/ds\right| $) followed by a negative one ($dz/ds=-\left| dz/ds\right| $).
These contributions cancel in pairs. In order that the contributions of two
consecutive points $P_{i}$ and $P_{i+1}$ are of the same (positive) sign, it
must occur that the portion of $C_{1}$ joining $P_{i}$ and $P_{i+1}$ winds
once around $C_{2}$. At the end, what we obtain as a result of the sum $%
\sum_{i=1}^{N}(\pm1)$ is the number $n$ of times that $C_{1}$ winds around $%
C_{2}$. Hence, we may use that $\int_{S}\mathbf{v}\cdot\mathbf{n}dS=n$ in
equations (\ref{17}) and (\ref{flux}) to obtain Gauss' result.

\end{document}